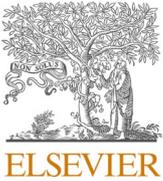

Contents lists available at ScienceDirect

# Measurement

journal homepage: www.elsevier.com/locate/measurement

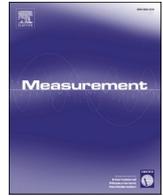

# ANN-based position and speed sensorless estimation for BLDC motors

Jose-Carlos Gamazo-Real [a, *], Víctor Martínez-Martínez [b, c], Jaime Gomez-Gil [b]

[a] Departamento de Sistemas Informáticos, Universidad Politécnica de Madrid, 28031 Madrid, Spain
[b] Departamento de Teoría de la Señal, Comunicaciones e Ingeniería Telemática, Universidad de Valladolid, Valladolid, Spain
[c] Facultad de Ciencias y Tecnología, Universidad Internacional Isabel I de Castilla, Burgos, Spain

ARTICLE INFO

ABSTRACT

BLDC motor applications require precise position and speed measurements, traditionally obtained with sensors. This article presents a method for estimating those measurements without position sensors using terminal phase voltages with attenuated spurious, acquired with a FPGA that also operates a PWM-controlled inverter. Voltages are labelled with electrical and virtual rotor states using an encoder that provides training and testing data for two three-layer ANNs with perceptron-based cascade topology. The first ANN estimates the position from features of voltages with incremental timestamps, and the second ANN estimates the speed from features of position differentials considering timestamps in an acquisition window. Sensor-based training and sensorless testing at 125–1,500 rpm with a loaded 8-pole-pair motor obtained mean absolute errors of 0.8 electrical degrees and 22 rpm. Results conclude that the overall position estimation significantly improved conventional and advanced methods, and the speed estimation slightly improved conventional methods, but was worse than in advanced ones.

## 1. Introduction

For over the last few years, brushless DC (BLDC) motors have quickly gained popularity due to their excellent characteristics and structure, which make them suitable for a wide variety of applications from low speed to high speed, such as electric vehicles and safety critical systems for the aerospace industry [1]. However, brushed DC (BDC) motors with position sensors have traditionally been used in applications that require precise detection (on the order of millimetres or centimetres), such as bio-inspired robots for medical cancer prevention [2] and robots for the automatic characterization of the radiation pattern of mobile phone base stations [3,4]. The advantages of BLDC motors over BDC, induction, and switched reluctance motors have promoted their popularity, such as exceptional speed and torque performance, energy savings, long operating life with reliability, and noiseless operation [5].

Many industrial applications with BLDC motors require precise speed and position measurements for rotor phase commutation. This operation is traditionally performed with sensors that are mounted inside the motor, as hall-effect units, or externally connected to the shaft, such as resolvers and encoders [6]. However, these devices increase the cost and size of the motors, restricting their applications, and the stressful operating conditions to which they are subjected could deteriorate signal measurements and cause failures [7]. To overcome these disadvantages, sensorless methods have emerged in recent years to estimate the rotor position and speed information from electrical signals, such as the phase terminal voltages [8], back-electromotive forces (BEMF) [9], conducting interval detection of inverter free-wheeling diodes [10], and current variation with stator core magnetic saturation [11].

Many efforts of different stakeholders have promoted the development of a wide variety of sensorless methods. The most popular category is based on the measurement of BEMF signals with direct or indirect techniques. In direct techniques, the BEMF zero-crossing in the floating phase is typically detected [12], but requires the use of filters that add attenuation and delay, which is avoided with signal conditioning in indirect techniques, such as the Third Harmonic Voltage Sensing [13]. However, the limited performance at low speed and the need for an open-loop starting strategy of BEMF-based methods have contributed to increase the use of techniques based on models and estimators, such as

---

the Sliding Mode Observer (SMO), the Extended Kalman Filter (EKF) and artificial intelligence (AI) algorithms. These techniques are widely applied to nonlinear systems, but depending on the technique, they have disadvantages such as the practical restrictions of SMO [14], and benefits such as accurate estimation with rapid convergence of EKF [15].

The results of the investigation included in the present work are of interest in the category of AI methods that are used for the identification and control of nonlinear dynamic systems, such as motor drives. Many publications of the literature are based on the use of AI techniques for motion control, as Vas [16] summarises. This category includes a variety of branches, such as Artificial Neural Networks (ANNs), Expert Systems, Fuzzy Logic (FZL), and Genetic Algorithms [17]. Among all branches of AI, ANNs seem to have the most relevant impact in the area of motor drives due to their ability to learn complex nonlinear functions with accuracy, fault tolerance, and flexibility [18]. ANNs are highly versatile as they can be combined with other complex algorithms, such as FZL to handle nonstatistical uncertain events [19], or they can use simple perceptron-based learning algorithms with low computational requirements, typically achieving similar performance to commonly accepted techniques such as EKF [20]. Furthermore, ANNs have been used for industrial applications with electric motors, such as estimating power consumption of BLDC motors in electrical vehicles such as drones [21] and identifying faults in motor elements, such as rotor and stator, using multilayer ANNs to optimise the operating life in BLDC motors [22,23] and induction motors through a perceptron-based topology [24]. The use of ANNs with a multilayer topology to build learning architectures also has wide application in estimating the position angle of rotational components such as shaft sensors [6] and developing position or speed control algorithms for BLDC motors, as a proportional integral derivative (PID), to improve the system dynamic response [25] and the stability under different motor load conditions through a deep-perceptron topology [26]. With these properties, ANNs represents a promising field of research for developing new position and speed estimation methods for BLDC motor drives [27].

The purpose of this article is to estimate the position and speed of a BLDC motor based solely on the phase terminal voltages without knowing the motor parameters and reference data from a sensor, by using simple processing resources. There are several estimation models that work with motor characteristics, such as the magnetization of the stator core [11] or the winding inductance variation [28], and estimators that perform complex functions on motor signals, such as SMO with a sinusoidal saturation function for BEMF estimation [29]. The use of ANNs is also extended for motor models other than BLDC, such as switched reluctance motors [30] and induction motors [31]. Furthermore, the use of ANNs for estimating is applied with different topologies, such as diagonally recurrent ANNs to obtain a rapid convergence [32], double ANN model connecting estimation models of current and BEMF signals [33], and feed-forward ANNs trained with the Levenberg-Marquardt algorithm as a simplified network topology for real-time deployment [34]. The ANNs implementation tends to be conducted in hardware-based platforms, such as Field-programmable Gate Arrays (FPGA), to develop innovative learning algorithms with high-level descriptions that enable comprehensive simulations and technology demonstrators of low-power consumption and reduced size [35], such as efficient and accurate real-time simulators for the fault diagnosis of electric drive systems in which the control algorithms and operating life forecast are of great relevance [36]. For this reason, it is expected that an ANN with a simple topology deployed in a FPGA would estimate the position and speed from BLDC motor signals with enough precision and low computational cost.

With this approach, an ANN-based method with multilayer perceptron (MLP) topology is presented in this article, and a series of experiments are fulfilled to determine its validity. To build the model, a FPGA-based test bench drives a loaded BLDC motor from standstill (in open loop) to variable speed operation (in closed loop) using a three-phase inverter controlled with pulse width modulated (PWM) signals. The terminal phase voltages of the motor are acquired with signal conditioning circuitry based on amplifiers and filters, and A/D converters built in the FPGA board. For rotor position detection, the typical six steps of the electrical commutation sequence are considered, as well as six virtual steps assigned to phase transitions that provide 12 rotor states in total. For ANNs training, the Backpropagation algorithm is used to obtain an optimal set of network parameters. An external encoder provides reference data to label the estimations (hypotheses) in ANNs training and to compare the estimations with the reference in method validation and testing. In sensorless mode, experimental tests are intensively performed at specific speeds in a wide range of 125–1,500 rpm.

The remainder of this article is structured as follows. In Section 2, the test bench to conduct the experiments is explained, followed by the description of the specific designs for motor signal conditioning and motor driving. Then, in Section 3, a novel estimation method based on ANNs with MLP topologies is developed, and the ANNs learning process is analysed considering the operation of the motor. In Section 4, the effectiveness of the method is validated by experiments with reference data from a position sensor, and the results are evaluated and discussed in comparison to other sensorless methods from related research. Finally, the conclusions are drawn in Section 5.

## 2. Experimental setup and data acquisition

In this section, an overview of the test bench used in the experimental phase is provided. The implementation of the motor signal conditioning and motor driving operations are analysed in detail as the main components to collect data for training and testing of the proposed method.

### 2.1. Test bench

The assessment of the method was carried out by conducting several experiments on a generic test bench that drove the motor and collected data. The test bench consisted of the processing hardware and learning software to support the test conditions of a real motor with an external load. Not only the test bench supported a Maxon EC 45 motor with the parameters shown in Table 1, but also delta and wye winding motors of different characteristics, such as voltage rating, load torque, and number of pole pairs.

The schematic layout of the experimental setup is depicted in Fig. 1. The motor driver was composed of a three-phase inverter and conditioning circuitry for amplification and filtering of motor voltages. Motor control and signal processing were relied on a Xilinx Spartan-6 FPGA and a real-time processor to generate PWM control signals, acquire conditioned motor voltages, and read encoder pulses. The reference positions were measured with a Kübler 2400 incremental encoder providing a resolution of 0.35 mechanical degrees with 1,024 pulses per revolution. Encoder data and motor voltages were sampled on separate A/D channels on the FPGA board. An illustration of the test bench is shown in Fig. 2.

**Table 1**
Parameters of the Maxon EC 45 motor.

| Description | Parameter | Value | Unit |
|---|---|---|---|
| Rated voltage | $V_{DC}$ | 12 | V |
| Rated power | $P_W$ | 30 | W |
| Phases | $p$ | 3 | – |
| Pole pairs | $K_p$ | 8 | – |
| Rated current | $I$ | 1.96 | A |
| Rated torque | $T$ | 53.2 | mNm |
| Phase resistance | $R$ | 1.4 | $\Omega$ |
| Phase inductance | $L$ | 0.56 | mH |
| Rotor inertia | $I_r$ | 92.5 | g·cm$^2$ |

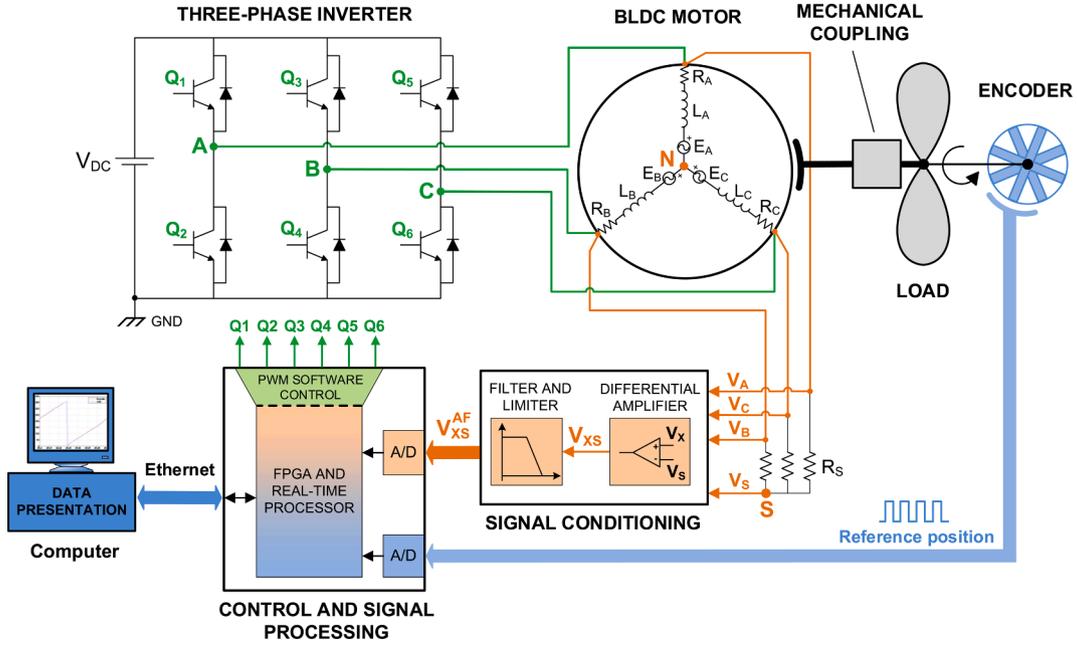

**Fig. 1.** Schematic layout of the experimental test bench.

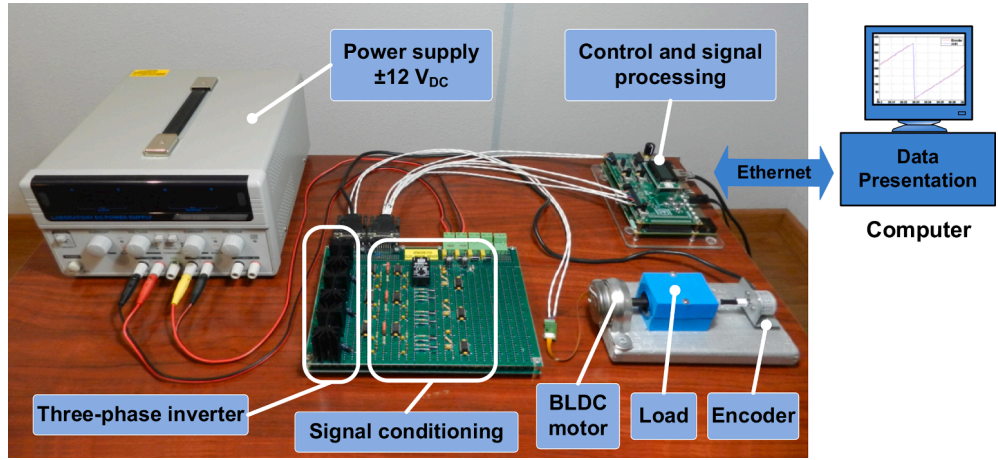

**Fig. 2.** Test bench used in the experimental phase.

## 2.2. Motor signals conditioning

The conditioning stage played a notable role in providing motor phase voltages within the appropriate ranges and with attenuated spurious components to be properly acquired by FPGA board A/D converters. This conditioning was based on differential amplification and filtering of motor signals over a virtual neutral point, as discussed in detail below.

The schematic layout of Fig. 1 includes a wye winding BLDC motor with a reference neutral point $N$. According to the mathematical model of a BLDC motor [37], the terminal phase voltages of three stator windings are given in matrix and reduced form as follows:

$$\begin{bmatrix} V_A \\ V_B \\ V_C \end{bmatrix} = \begin{bmatrix} R_A & 0 & 0 \\ 0 & R_B & 0 \\ 0 & 0 & R_C \end{bmatrix} \bullet \begin{bmatrix} I_A \\ I_B \\ I_C \end{bmatrix} + \begin{bmatrix} L_A & -M_{AB} & -M_{AC} \\ -M_{AB} & L_B & -M_{BC} \\ -M_{AC} & -M_{BC} & L_C \end{bmatrix} \bullet \frac{d}{dt} \begin{bmatrix} I_A \\ I_B \\ I_C \end{bmatrix} + \begin{bmatrix} E_A \\ E_B \\ E_C \end{bmatrix} \tag{1}$$

$$V_X = R \bullet I_X + (L - M) \bullet \frac{d}{dt} I_X + E_X \tag{2}$$

where $X$ is the motor phase ($A$, $B$, and $C$), $R$ is the stator resistance (assumed to be similar for all windings), $I_X$ is the armature current, $L$ and $M$ are, respectively, the self-inductance and the mutual inductance (assumed both constant and similar for all windings), and $E_X$ is the trapezoidal-shaped BEMF.

To perform phase voltage measurements with low commutation noise, a virtual neutral point $S$ with three resistors in wye configuration was used. The conditioning circuit designed for each motor phase $X$ is illustrated in Fig. 3. The common-mode noise at the virtual point was reduced with differential amplifiers of high common-mode rejection and low noise, such as the INA126 model with a gain of 5 V/V. The terminal phase voltages with respect to the virtual point are:

$$V_{XS} = V_{XN} + V_{NS} = V_X - V_N + V_N - V_S = V_X - V_S \tag{3}$$

As the operation of the PWM-based three-phase inverter presented high frequency components, they were attenuated with two filters [38]: a restrictive low-pass filter ($R_a$, $C_a$) for frequencies <20 kHz and a low-pass Pi filter ($L_b$, $C_b$) for high-frequency spurious components from 100 to 200 kHz. In the final stage of the circuit, a current limiting resistor and clamping Schottky diodes were included to protect the next stage of data processing against overcurrent, undervoltage and overvoltage. Motors of different ratings, such as 12 or 24 V nominal voltage, were supported as the input conditioning supply voltage $V_{DC}$ was configured to that nominal limit. However, the output conditioning levels were adjusted to the $V_{CC}$ supply voltage of the input interface on the processing board, as a range of ± 5 V and 10 mA per analog input. After the conditioning stage, the high frequency components of the conditioned voltages $V_{XS}^{AF}$ were attenuated without any relevant influence on the dynamic performance, such as slew rate and speed response. The acquisition of the conditioned voltages was carried out in the processing board with A/D converters of 16 bits and a sampling frequency of 100 kHz (period of 10 μs). As an example, the actual conditioned voltages of phase *A* are shown in Fig. 4.

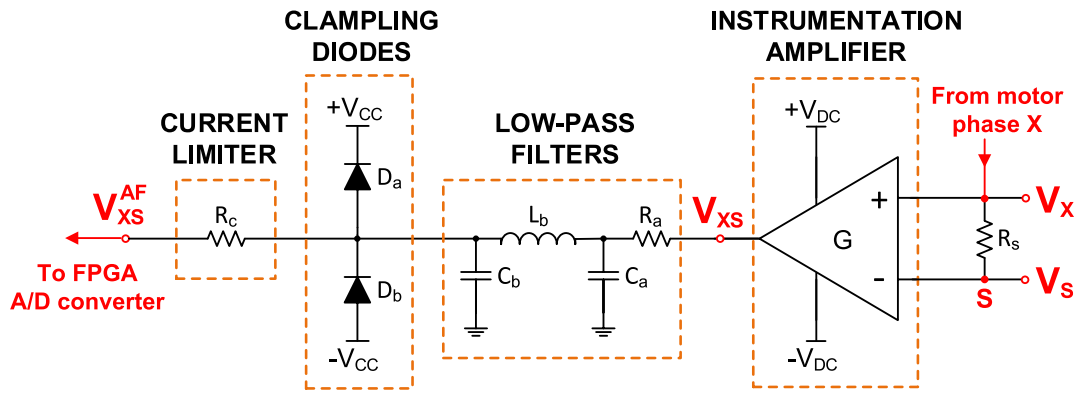

**Fig. 3.** Stages of the signal conditioning circuit for each motor phase *X*.

### 2.3. Motor driving

The main components of the motor driver were the commutation devices of the three-phase inverter and the generation of their gate commutation signals. The power transistors *Q1* to *Q6* of the inverter, such as the UC2950T half-bridge model, were switched on and off with PWM signals according to a six-step electrical sequence. This sequence excited only two of the three phases and the third phase remained floating at any instant in time [39], as presented in Table 2. The PWM signals were generated by a FPGA-based control board to switch the transistors with a fixed duty cycle ($D = T_{ON}/T_{PWM}$) and a variable commutation period between phases to change the motor speed. As the commutation properties of the transistors (turn-on and turn-off times close to 200 ns) limited the PWM duty cycle, it was adjusted to the minimum commutation period of these devices through $T_{ON}$ (time that the PWM signal remained at a high level). This strategy provided the flexibility to use motors of different nominal voltages by limiting the proportion of PWM duty cycle used, obtaining a driver for low noise operation at high speeds and controlled with PWM configurations adapted to the motor characteristics [40]. The driver was operated with a 20 kHz PWM frequency and 50% duty cycle after various evaluation tests.

**Table 2**
Six-step sequence of clockwise BLDC motor rotation and relation to rotor position.

| Sequence number | Active transistors | | Active phases | | | Rotor position | |
|---|---|---|---|---|---|---|---|
| | High | Low | A | B | C | Electrical degree | Mechanical degree |
| 1 | Q1 | Q4 | On | On | Off | 0–60 | $(0–60)/K_p$ |
| 2 | Q1 | Q6 | On | Off | On | 60–120 | $(60–120)/K_p$ |
| 3 | Q3 | Q6 | Off | On | On | 120–180 | $(120–180)/K_p$ |
| 4 | Q3 | Q2 | On | On | Off | 180–240 | $(180–240)/K_p$ |
| 5 | Q5 | Q2 | On | Off | On | 240–300 | $(240–300)/K_p$ |
| 6 | Q5 | Q4 | Off | On | On | 300–360 | $(300–360)/K_p$ |

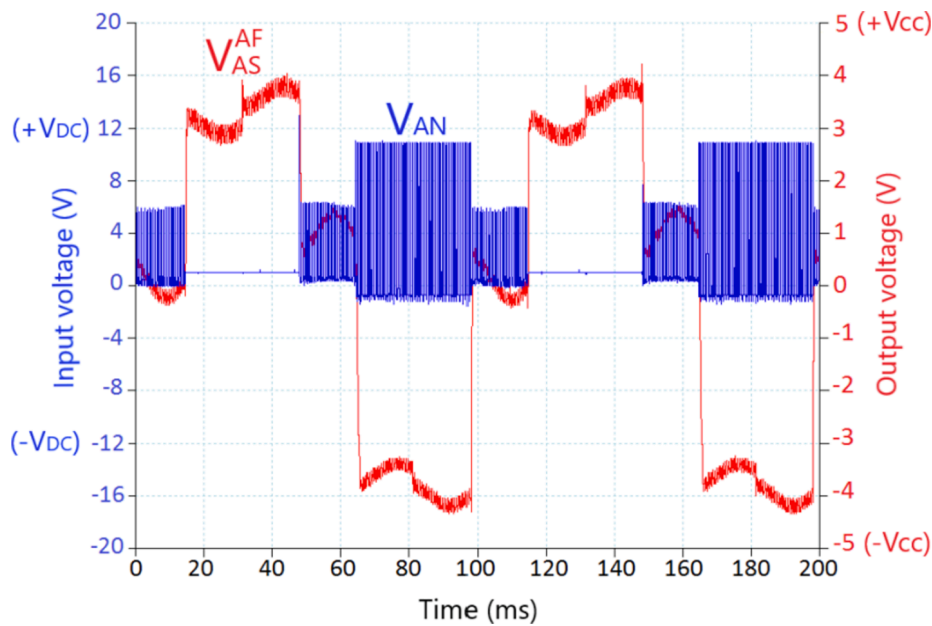

**Fig. 4.** Actual motor voltages for phase *A* at the input ($V_{AN}$) and output ($V_{AS}^{AF}$) of the conditioning circuit.

As a three-phase inverter was used to perform the commutation of the stator windings through the six-step sequence, at each step the motor phase had its upper or lower leg in a disconnected state to switch only two transistors at the same time [10]. This operation represented the basis to estimate six position steps in an electrical cycle, which are related to the mechanical cycle as follows:

$$T_m = K_p \cdot T_e \tag{4}$$

where $T_m$ is the mechanical period, $T_e$ is the electrical period, and $K_p$ is the number of rotor-pole pairs.

Based on this operation, the starting of the motor from standstill was performed with an open-loop ramp-up model without rotor pre-alignment using the PWM signals for transistor switching. The motor was linearly accelerated to a reference speed, which simplified the starting and considerably avoided traditional reverse rotational problems [41]. It provided a stable starting alternative with high torque and low start-up time without the drawbacks of conventional techniques, such as reduced motor signal levels and signal-to-noise ratio at low speeds that often influence on BEMF detection [42]. Once the motor reached a speed close to 100 rpm, a closed-loop control was used with the encoder data as the reference position.

## 3. Experimental method for position and speed estimation

This section describes the proposed method for estimating position and speed using information from motor voltages only. It is based on MLP-ANN topologies which estimate the motor rotational angle as the main parameter and then obtains the motor speed. The proposal represents a generic estimation method without considering the motor model or the detailed analysis of its electrical signals. Obtaining the Virtual Sequence Numbers (VSN) is described first, as it is one of the basis of the method. Next, the estimation algorithm overview is explained to achieve a broad perspective of the proposal. Subsequently, the proposed MLP-ANN topologies and the process of ANNs training, validation, and testing are described in detail.

### 3.1. Virtual sequence numbers

In a BLDC motor, the position detection is related to the electrical rotor position and the commutation sequence number. Furthermore, in these motors, the number of poles limits the resolution of the rotor angle, so that only a certain number of rotor steps are possible and the position detection methods are limited by this characteristic. To facilitate the identification of the real rotational angle, the proposed method considered six virtual steps corresponding to the transitions between the consecutive electrical steps of the rotor that are shown in Table 2. In total, 12 rotor steps (6 real steps and 6 virtual steps) were provided in each electrical cycle, which were called VSN. Each of them represented an angular range per electrical cycle with two components (sine and cosine) and were identified by the angle projections on the unit circle, as presented in Table 3. The use of these components is a common approach to increase the ANNs performance for small angle variations when the input changes are significantly low [43].

VSNs contributed to reduce the influence of noise over ANNs training because input phase voltages were mapped to a discrete output VSN label. Each label represented an angular range rather than a continuous variable, such as the rotational angle, which is more prone to noise distortions. Furthermore, VSNs represented an increase in the number of outputs for ANNs training up to 96 labels (12 VSN multiplied by 8 pole pairs for the EC 45 motor) and the minimum measurable angle was from 7.5 to 3.75 mechanical degrees. It was a simple way to increase the angle resolution because, according to Table 2, in a sequence transition, only two of the three phases changed in each sequence step. That condition for detecting a transition was considered easy to learn in ANNs, since the sign of two phases changed simultaneously and near the sign change the voltage levels were close to zero, which was similar to a zero-crossing detection (ZCD) operation. Using this approach, the voltage inputs from each training example were mapped to a sine–cosine pair as an output label to identify the real position components (obtained from encoder reference positions) that the ANN algorithm should learn. In each VSN, the mean value of the sine and cosine ranges was used as a label to reduce the detection error by half. As an example, in the VSN 2 corresponding to an electrical transition, the sine and cosine components were 0.098 and 0.995, respectively, and the mechanical angles, read from the encoder and located in the same range of the components, were assigned the same mean value for training.

**Table 3**
Equivalence between VSNs and mechanical rotor positions.

| VSN | Electrical step with sequence number or electrical transition | Rotor Position (mechanical degree) | | | |
|---|---|---|---|---|---|
| | | Generic Angle | Position in 1st electrical cycle for Maxon EC 45 motor ($K_p = 8$) | | |
| | | | Angle | Sine component | Cosine component |
| 1 | 1 | $(0-30)/K_p$ | 0–3.75 | 0–0.065 | 1–0.998 |
| 2 | 1 → 2 | $(30-60)/K_p$ | 3.75–7.5 | 0.065–0.131 | 0.998–0.991 |
| 3 | 2 | $(60-90)/K_p$ | 7.5–11.25 | 0.131–0.195 | 0.991–0.981 |
| 4 | 2 → 3 | $(90-120)/K_p$ | 11.25–15 | 0.195–0.259 | 0.981–0.966 |
| 5 | 3 | $(120-150)/K_p$ | 15–18.75 | 0.259–0.321 | 0.966–0.947 |
| 6 | 3 → 4 | $(150-180)/K_p$ | 18.75–22.5 | 0.321–0.383 | 0.947–0.924 |
| 7 | 4 | $(180-210)/K_p$ | 22.5–26.25 | 0.383–0.383 | 0.947–0.924 |
| 8 | 4 → 5 | $(210-240)/K_p$ | 26.25–30 | 0.383–0.5 | 0.924–0.866 |
| 9 | 5 | $(240-270)/K_p$ | 30–33.75 | 0.5–0.556 | 0.866–0.831 |
| 10 | 5 → 6 | $(270-300)/K_p$ | 33.75–37.5 | 0.556–0.609 | 0.831–0.793 |
| 11 | 6 | $(300-330)/K_p$ | 37.5–41.25 | 0.609–0.659 | 0.793–0.752 |
| 12 | 6 → 1 | $(330-360)/K_p$ | 41.25–45 | 0.659–0.707 | 0.752–0.707 |

The signal processing circuit acquired the encoder reference positions and assigned those positions to VSNs taking into account the corresponding angle ranges considered in Table 3, which depended on $K_p$ for a specific motor. Fig. 5 shows the graphs of three motor voltages after signal conditioning ($V_{AS}^{AF}$, $V_{BS}^{AF}$, and $V_{CS}^{AF}$) and their relationship to 96 VSN labels per mechanical cycle (12 VSN per electrical cycle) for the EC 45 motor at a speed of 1,500 rpm. These graphs illustrate the mapping between voltage levels and rotor position that the ANNs performed after training to estimate position and speed. In the first electrical cycle of Fig. 5, it is observed that odd VSNs correspond to conventional sequence numbers for commutating inverter transistors with one inactive phase and two active phases, which provide stable voltages. However, even VSNs correspond to electrical transitions of phase voltages with one phase active and two phases changing sign, equivalent to a double zero-crossing operation.

As an illustrative example of the VSN mapping process for training, the position steps VSN 7 and VSN 12 of the first electrical cycle and the phase voltage levels for each step are highlighted in Fig. 5. For VSN 7, the points are $a1 \approx$ -4 V and $b1 \approx$ 4 V ($A$ and $B$ are the active phases), and $c1 \approx 0$ V ($C$ is the inactive phase). For VSN 12, the points are $b2 \approx$ -4 V ($B$ is the active phase), while $a2$ and $c2$ are for phases in transition ($A$ and $C$) whose voltages correspond to level changes, similar to a double zero-crossing operation (change from 0 V to $\pm 4$ V, and vice versa). These values are representative of the input data (voltage levels) and the output labelling data (VSNs) that the ANNs used in their learning

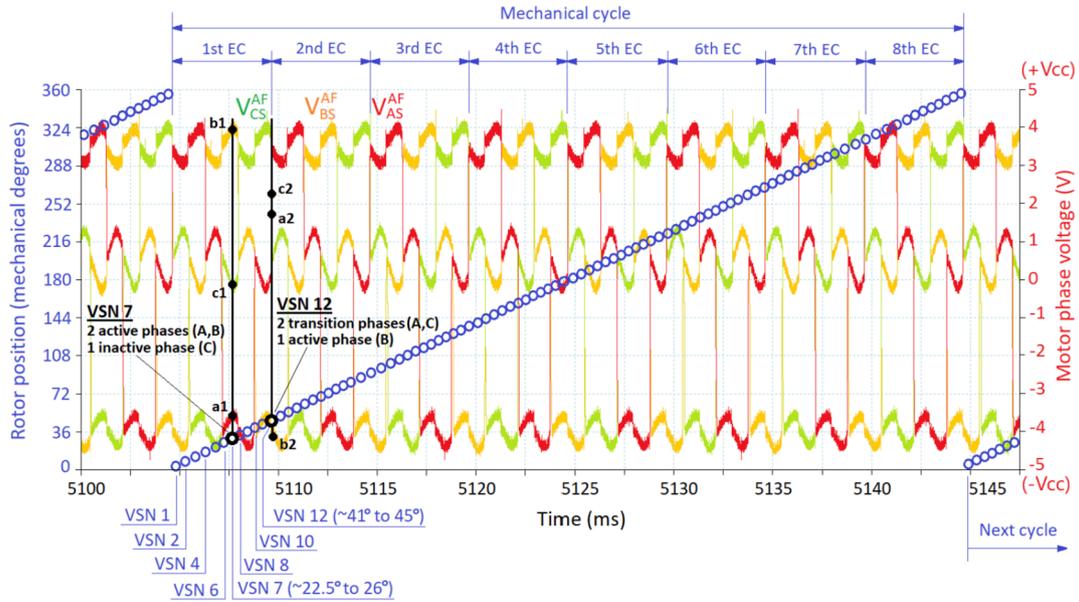

**Fig. 5.** Three-phase motor voltage levels after signal conditioning ($V_{XS}^{AF}$) related to VSN labels (drawn in blue circles) in each electrical cycle (EC) and mechanical rotor positions for Maxon EC 45 motor ($K_p = 8$) at a speed of 1,500 rpm. (For interpretation of the references to colour in this figure legend, the reader is referred to the web version of this article.)

process at each sequence step to provide the position estimation.

### 3.2. ANN-based estimation algorithm

The proposed method was developed with multilayer ANNs. The fundamental part of the method estimated the rotor position and the rotor speed estimation was based on the position results, as shown in Fig. 6. When selecting the type and topology of ANNs, the simplicity and the use of low computational resources were considered as the main characteristics to enable their implementation in low-cost hardware devices for online training or on a computer for offline training. An MLP topology with fully connected layers was selected instead of more complex architectures, such as convolutional or recurrent ANNs, which are typically used in image-based methods for machine diagnosis that requires more computing resources [44]. The selected topology allowed for prototyping and implementation in commercial embedded devices such as FPGAs.

The MLP topology was implemented with three layers. In the input layer, the number of nodes was related to the input variables to be processed (motor voltages and acquisition time), while in the output layer the number of nodes depended on the hypothesis function. In the hidden layer, several structures with different number of nodes were independently developed and trained to select the topology with the maximum success rate in the validation dataset. The tan-sigmoid activation function $g_1(x) = -1 + 2/(1 + exp(-2 \cdot x).)$ was used in the hidden layer nodes to establish nonlinear relations among layer inputs and outputs with a limited range to ±1. However, the linear activation function $g_2(x) = x$ was selected for the output layer nodes to allow unlimited values. It was admitted that a network with tan-sigmoid hidden neurons and linear output neurons can adapt quite well to multidimensional mapping problems when fed with consistent data and the hidden layer has enough neurons [34].

#### 3.2.1. Position estimation ANN

The position estimation MLP-ANN topology consisted of 10 input nodes, five hidden nodes, and two output nodes, as illustrated in Fig. 7. The ANN inputs provided information on the three-phase motor signals in each acquisition time slot and their evolution over time. The inputs were the terminal phase voltages at the current acquisition time instant ($V_{XS}^{AF}(t)$), the terminal phase voltages at the next acquisition time instant ($V_{XS}^{AF}(t + \Delta t)$), and the corresponding acquisition time slot as time difference ($\Delta t$). To obtain additional information on voltage changes, such as the sign of the signals, a set of three *voltMul* features was synthesised as the product of the terminal phase voltages in the current and previous time instants (*voltMul$_A$*, *voltMul$_B$*, and *voltMul$_C$*), as shown in Fig. 6. The new position features were calculated as follows:

$$voltMul^t_X = V^{AF}_{XS}(t) \cdot V^{AF}_{XS}(t + \Delta t) \tag{5}$$

where *X* is the motor phase (*A*, *B*, and *C*) and $\Delta t$ is the time slot in which the product operation is carried out to obtain the feature.

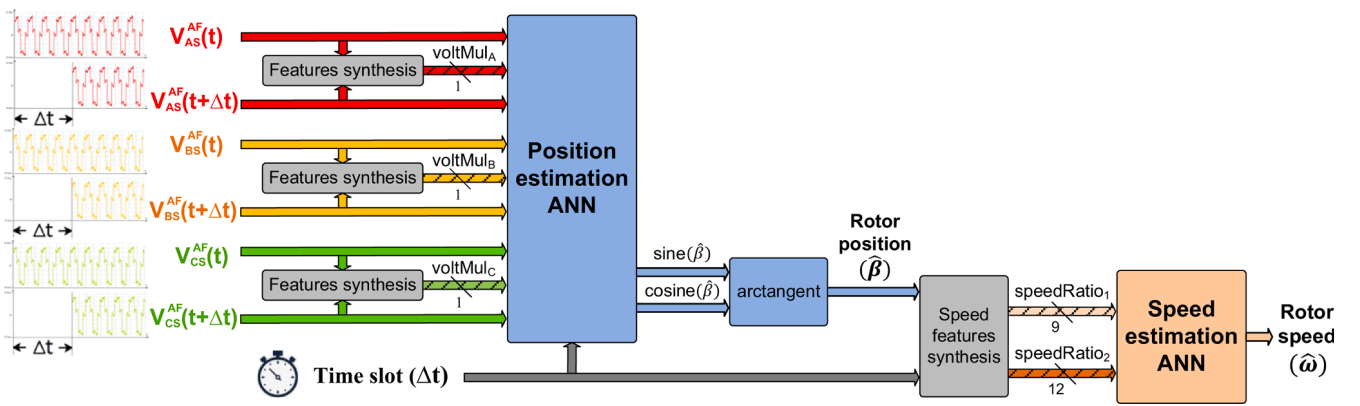

**Fig. 6.** Block diagram of the ANN-based estimation algorithm.

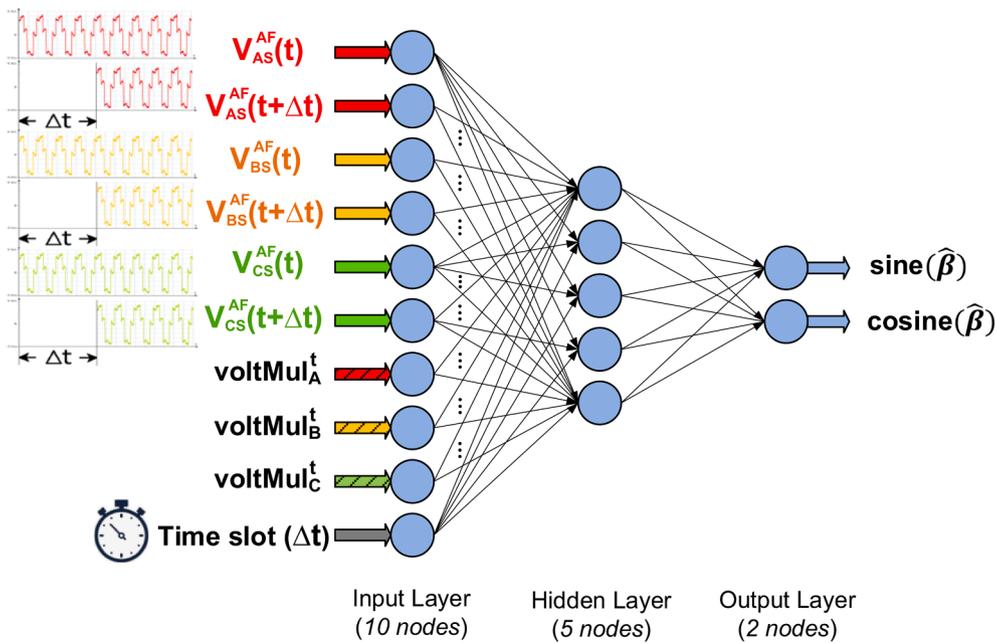

**Fig. 7.** Position estimation ANN with a three-layer MLP topology.

The input to output mapping was provided as the sine and cosine components of the estimated rotational angle ($\widehat{\beta}$) rather than the estimated angle itself. Finally, the estimated position was the result of the arctangent function, which also avoided the problem of inconsistent values such as those greater than one.

### 3.2.2. Speed estimation ANN

The speed estimation was also performed with a three-layer MLP-ANN topology. However, it could have been implemented as a simple conventional observer based on the cumulative sum of the number of mechanical rotor cycles per unit time. Based solely on this simple mathematical algorithm, the resolution would have depended on the position estimation errors and the correct identification of the rotor cycle. In that case, the cycle detection would have been more sensitive to position deviations as ANNs perform additional data filtering tasks. In the proposed ANN, the speed resolution was also based on position errors, but the identification of the cycle over time only used the features synthesised in the ANN training.

The speed estimation was developed with two stages, as shown in Fig. 6. The first stage calculated a set of *speedRatio* features as a function of the rotor positions and their timestamps, and the second stage implemented the ANN taking the speed ratio features as input, as depicted in Fig. 8. Two *speedRatio* features were obtained based on position differentials and their associated timestamps. The values of the *speedRatio1* feature were calculated with the difference of each position sample and the last sample in an acquisition window. It provided features in incrementally sized time slots to detect instantaneous speed variations and unexpected deviations (outliers). The values of the *speedRatio2* feature considered position samples in consecutive rotor cycles with the same VSN, which provided an observer of the position changes during complete cycles to detect periodic variations in speed. The speed ratio features are given as follows:

$$speedRatio_1^p = \frac{\widehat{\beta}_n - \widehat{\beta}_{n-p}}{t_n - t_{n-p}} \tag{6}$$

$$speedRatio_2^q = \frac{\widehat{\beta}_k^q - \widehat{\beta}_{k-1}^q}{t_k^q - t_{k-1}^q} \tag{7}$$

where $\Delta t = t_i - t_{i\cdot j}$ is a time slot with $t_i$ as the acquisition timestamp of the position sample $i$ that corresponds to the rotor position $\widehat{\beta}_i$, $n$ is the size of the acquisition window, $p$ is the index of a position sample in the window ($p = 1$ to $n-1$), $k$ and $k-1$ are the indexes of position samples with the same VSN of value $q$ ($q = 1-12$) in two consecutive rotor cycles.

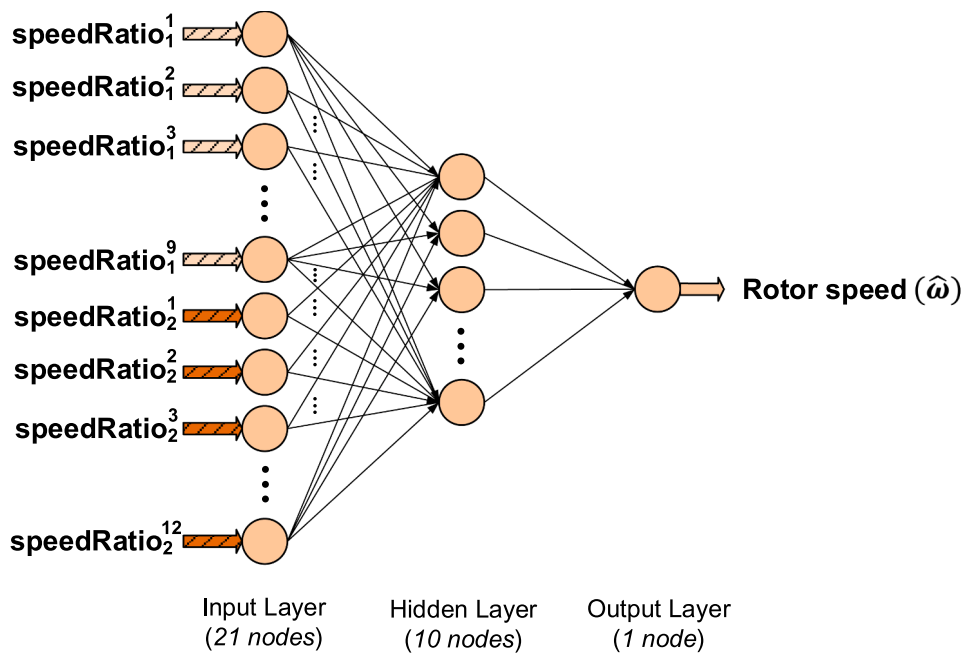

**Fig. 8.** Speed estimation ANN with a three-layer MLP topology.

Using this approach, the values of the *speedRatio1* feature were collected with a window of size $n = 10$ that provided nine speed ratio values. For the *speedRatio2* feature, 12 speed ratio values were obtained since each VSN provided a rotor position and, consequently, its associated speed ratio. Therefore, 21 nodes were used in the input layer. The hidden and output layers were implemented with ten nodes and one node, respectively, after various evaluation tests.

### 3.3. Motor operation and ANNs learning

The measurement of BLDC motor and encoder signals was used for ANNs training and further testing. The learning process consisted of several stages and required operation of the motor in sensor-based and sensorless modes, as illustrated in Fig. 9.

In the first stage, the motor was started from standstill with an open-loop ramp-up control to develop speed. It provided a simple start-up strategy over conventional BEMF based techniques, which are more typical but not applicable for self-control from standstill because BEMF signals are very low in this state. Sensor-based control was then enabled at a speed close to 125 rpm and the encoder signal was used as the reference position to control the drive. At this stage, during variable speed operation at 125–1,500 rpm, the motor phase voltages and encoder data were measured and stored as training examples for the position estimation ANN. Based on those examples and the position estimations, the speed ratio features were obtained to train the speed estimation ANN. In the last stage of the process, the motor was driving in sensorless mode to evaluate the performance of the proposed method with respect to tracking the encoder reference signal.

As the main operation of the proposed method, the training of the position estimation ANN was performed with the Backpropagation algorithm once $m$ examples ($x^{(i)}, y^{(i)}$) were acquired, where $i = 1$ to $m$, $x = x^{(i)}$ is the input vector of motor phase voltages and timestamps, and $y = y^{(i)}$ is the output vector of VSN labels associated to encoder positions. The weights of each ANN node were randomly initialized and the acquired motor data $x$ was used as ANN input data ($a^{(1)} = x$). Next, in the forward propagation phase of the algorithm, the vectors of the remaining neurons $a^{(j)}$ were calculated, where $j = 2$ (hidden layer) and $j = 3$ (output layer). The ANN hypothesis output vector $h_{\Theta}(x)$ was calculated as expressed below:

$$h_{\Theta}(x) = a^{(3)} = g_2\left(\Theta^{(2)} \bullet a^{(2)}\right) \tag{8}$$

$$a^{(2)} = g_1\left(\Theta^{(1)} \bullet a^{(1)}\right) \tag{9}$$

where $g_1$ is the tan-sigmoid activation function for the hidden layer, $g_2$ is the linear activation function for the output layer, $\Theta^{(1)}$ is the weight matrix to control the mapping of the input layer to the hidden layer, and $\Theta^{(2)}$ is the weight matrix to control the mapping of the hidden layer to the output layer.

In the back propagation phase of the algorithm, the cost errors $\delta^{(j)}$ were obtained and the cost was minimised with an optimal set of parameters $\Theta^{(j)}$, considered as the network weights. For the output layer ($j = 3$), the error vector was the difference of the actual results and the correct output. However, for the hidden layer ($j = 2$), the error vector was calculated with the product of the output layer error and the hidden-output weight matrix, and the derivative of the activation function for the hidden layer ($g_1$') applied to the input-hidden transform:

$$\delta^{(3)} = a^{(3)} - y \tag{10}$$

$$\delta^{(2)} = \left((\Theta^{(2)})^T \bullet \delta^{(3)}\right) * g_1{}'\left(\Theta^{(1)} \bullet a^{(1)}\right) \tag{11}$$

where the operation "*" represents an element-wise multiplication of vectors and matrices, and the superscript "T" indicates a transposed matrix.

The procedures of ANN training, validation, and testing were performed on measurements of triangle and up-down variable motor speeds over the range of 125–1,500 rpm, which covered the main motor

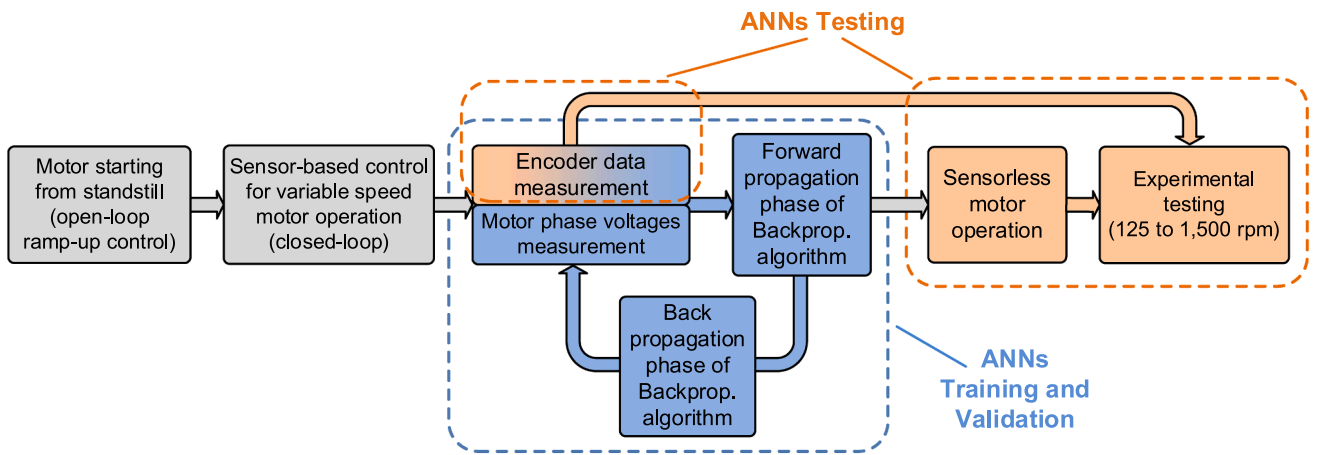

**Fig. 9.** Process of motor operation and data measurement for ANNs learning.

speed–torque operating area, as described in Table 4. The dataset measurements were collected at a sampling rate of 100 kHz and rearranged randomly prior to allocation to a specific subset. It provided symmetry breaking and generalization error reduction that substantially avoided the influence of overfitting (high variance) and underfitting (high bias) problems. In this work, the total number of ground truth measured data for training, validation, and testing was equal to 343,500, which may contain outliers and missing values than may skew the results. Because proper data division is relevant as the results may be biased depending on how the data is split, the measured values were divided into fractions. The data division was adopted as a 40% fraction of the ground truth data for training, a 10% fraction for cross-validation to evaluate ANNs generalization, and a 50% fraction for testing to provide an independent measure of ANNs performance after training. It should be noted that the training and cross-validation sets were selected small enough to reduce the processing load of the training process with an adequate learning rate, allowing the possibility to train offline on the computer or online on the FPGA board.

After training and testing the ANNs, the motor operation in sensorless mode started with the ANNs data and without the support of the encoder data. In the experimental tests over the full speed range, the position and speed results of the ANNs estimations were compared with the encoder data, as analysed in the next section.

## 4. Experimental results and discussion

The experiments were carried out to verify the performance of the position and speed estimations with the proposed method. The experimental test bench is shown in Figs. 1 and 2, in which the Maxon EC 45 motor was used (see parameters in Table 1) with a medium load torque of 35 mNm coupled to the shaft. The performance was analysed with the loaded BLDC motor and a shaft encoder in the speed range of 125–1,500 rpm.

### 4.1. Evaluation guidelines of estimation results

To evaluate the performance of the proposed method, not only the success of the training was considered, but also the criteria for a correct assessment of the estimation results with respect to the real target values. This was carried out with fractional parts with random ordering of ground truth data for training, validation and test sets, and the selection among other fitting models, such as Support Vector Machines, of a sufficiently simple and flexible learning topology. Likewise, as the statistical criteria played a relevant role in the evaluation, the F-score and the mean absolute error (MAE) were used as metrics. The F-score is the harmonic mean of precision and recall, and was used as a combined evaluation metric to assess average rates rather than the accuracy. However, the accuracy also provided an additional ponderation as evaluates the ratio of true cases (estimated correctly) and the total number of examples. The MAE quantifies the difference between the paired estimated and target values in a large set of examples, and was intended to minimise with an optimal set of ANN parameters $\boldsymbol{\Theta}^{(j)}$ as network weights. Their formulas are reminded next:

$$F - score = 2 \bullet \frac{P \bullet R}{P + R} \tag{12}$$

where $P$ is the precision as the ratio of true positives and the number of predicted positive cases (true and false positives), and $R$ is the recall as the ratio of true positives and the number of actual positive cases (true positives and false negatives).

$$MAE = \frac{1}{n} \sum_{i=1}^{n} |\widehat{y}_i - y_i| \tag{13}$$

where $n$ is the total number of examples, and $\widehat{y}$ and $\boldsymbol{y}$ are the estimated and target vectors, respectively. In the case of the rotor position as the main estimation result of the proposed method, $\boldsymbol{y}$ corresponds to the position reference from the encoder and $\widehat{y}$ have two components (sine and cosine) to obtain the estimated position as follows:

$$\widehat{\beta} = \arctan\left(\frac{\sin(\widehat{\beta})}{\cos(\widehat{\beta})}\right) \tag{14}$$

### 4.2. Performance of VSN estimation

The detection of the rotor positions was associated with the estimation of VSNs, so its performance evaluation was considered a key factor. As described in Table 4, the ANNs were trained, validated, and tested with two categories of datasets (triangle and up-down) on 343,500 motor measurements. However, the experimental testing of ANNs was performed on more than 1.75 million motor measurements (over 30,000 rotor mechanical cycles) at low, medium, high, and very high fixed speeds over the range of 125–1,500 rpm, as shown in Table 5.

The performance of the VSN estimation was evaluated as the ratio between the unknown states (no match with a known state), the successful states, and the failed states related to the actual position of the rotor. Fig. 10 shows the accuracy of 32 speed tests over the full range and indicates that 92% are successful states (true positives), 7.5% are unknown (false negatives and false positives), and 0.5% are erroneous (true negatives). The rate of unknown states increased significantly at low speeds (about 30%) due to the difficulty in BLDC motors to detect the rotor position near standstill. The proposed method significantly improved the performance when the motor speed was higher than 175 rpm due to the drastic reduction of unknown states, although the estimation between 125 and 175 rpm was obtained with a moderate error.

**Table 4**
Datasets for ANNs training, validation, and testing in the full speed range.

| Category of dataset | Motor speed range (rpm) | Description | Motor measurements ($x^{(i)}$, $y^{(i)}$) | Rotor mechanical cycles |
|---|---|---|---|---|
| Triangle | 85–950 | The speed increases in the range and then falls into a triangle shape. The speed varies in steps of 1–10 rpm, and the frequency of speed variation changes. | 159,350 | ≈ 3,187 |
| Up–down | 85–1,500 | The speed increases and falls alternately within the full range. Steps: 85 → 250 → 125 → 600 → 400 → 950 → 750 → 1,500 | 184,150 | ≈ 3,683 |

**Table 5**
Dataset for ANNs experimental testing at specific speeds over the full range.

| Category of dataset | Motor speed value (rpm) | Motor measurements ($x^{(i)}$, $y^{(i)}$) | Rotor mechanical cycles |
|---|---|---|---|
| Low speed | 125, 175, 200, 225, 250 | 157,630 | ≈ 2,866 |
| Medium speed | 325, 400, 475, 540, 600, 650, 700 | 567,215 | ≈ 10,313 |
| High speed | 725, 750, 800, 850, 900, 950 | 793,430 | ≈ 14,426 |
| Very high speed | 975, 1,000, 1,050, 1,100, 1,150, 1,200, 1,300, 1,400, 1,500 | 237,820 | ≈ 4,324 |

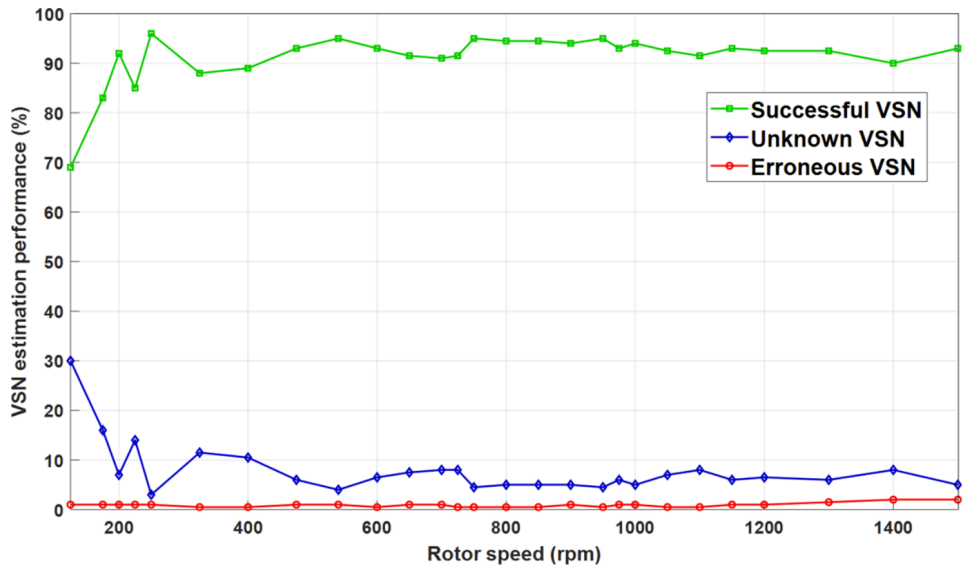

**Fig. 10.** Performance of VSN estimation related to successful, unknown, and erroneous motor position states in the speed range of 125–1,500 rpm.

As noted, the performance of the estimation remained high in the full speed range, but decreased at very low speeds. The operation close to standstill was not considered in the present work and was handled with an open-loop ramp model without rotor pre-alignment using PWM control, as explained in Section 2. To cope with these limitations, a different learning procedure may be specifically applied in these sensitive areas. Some relevant works are described in the literature for estimating the rotor position from standstill to a speed below 100 rpm. Examples of methods that consider those measurements are the Second-order generated Integrator [45], the Indirect High-frequency Signal Injection [46], and those based on the effect of magnetic saturation [47].

The performance measurements of the proposed ANN model for VSN multiclass classification were provided in the dataset described above, which was larger than the cross-validation set and was highly focused on specific speeds. The use of large datasets allowed the performance to be evaluated in detail and helped to verify the initial aim of reducing computational resources with simple ANN topologies and small training sets. The evaluation for each speed under test was determined with the values of the parameters based on the confusion matrix, such as accuracy, precision, recall and, mainly, F-score. However, only the overall values of those parameters were provided in this analysis to simplify, as they were all very similar for each speed. The F-score was preferentially considered as the combined evaluation metric, since the cost of false positives was similar to that of false negatives and a correct association of VSN labels with examples was more important. The results were 0.961 for the F-score and, for reference only, 0.925 for the accuracy, supporting that the classifier was extremely precise and accurate, and omitted a relatively small number of examples.

### 4.3. Performance of position and speed estimation

As mentioned, the core of the developed estimator was the position ANN-based algorithm and the performance of the method was based on the ability of the ANNs to estimate the VSNs. As illustrated in Fig. 10, the performance of VSN estimation was validated over the full speed range of 125–1,500 rpm. Furthermore, in this range, the tracking of the encoder measured data was compared with the estimations to obtain the MAE of position and speed.

The performance results at different motor speeds are summarized in Table 6. The results in bold type show a small position MAE below 0.8 electrical degrees and a moderately high relative speed error of 5% (MAE below 22 rpm) over the full speed range.

Performance results were obtained after detailed graphing and numerical analysis for each motor speed. Because the results were very similar for the speeds under test, only the analyses for 175 rpm (low speed, near to the starting) are included in Figs. 11 and 12, and the results for 850 rpm (high speed) are illustrated in Figs. 13 and 14. The position graphs only include one mechanical cycle (0–360°) to evaluate in detail the tracking of the encoder reference during a cycle and at the cycle transition. However, the speed graphs cover a large number of motor voltage measurement samples (three-phase voltages per sample) to validate the steady-state performance over various mechanical cycles. Each graph includes a zoom view to analyse the instantaneous errors of the estimated position and speed with respect to the encoder reference, which were used to obtain the MAE results for a specific speed test. Position overshoot or undershoot were not relevant near cycle transitions, and there was a slight fluctuation around reference in the position and speed graphs, which remains stable within a tolerance margin at stationary.

**Table 6**
Performance measurements of experimental position and speed tests.

| Motor Speed (rpm) | Speed category | Position MAE (mech. degree) | Position MAE (electrical degree) | Speed MAE (rpm) | Relative Speed Error (%) |
|---|---|---|---|---|---|
| 125 | Low | 6.1 | 0.76 | 5.4 | 4.3 |
| 175 | Low | 5.7 | 0.71 | 7.3 | 4.2 |
| 325 | Medium | 6.2 | 0.77 | 17.9 | 5.5 |
| 475 | Medium | 6.4 | 0.8 | 12.8 | 2.7 |
| 600 | Medium | 6.4 | 0.8 | 22.8 | 3.8 |
| 725 | High | 6.3 | 0.79 | 32.6 | 4.5 |
| 850–1,500 | High to very high (10 speed values) | ≈ 6.5 | ≈ 0.8 | 21.2–37.5 | ≈ 2.5 |
| **125–1,500** | **Low to very high (Full range)** | **< 6.5 mech. deg.** | **< 0.8 elect. deg.** | **< 22 rpm** | **< 5%** |

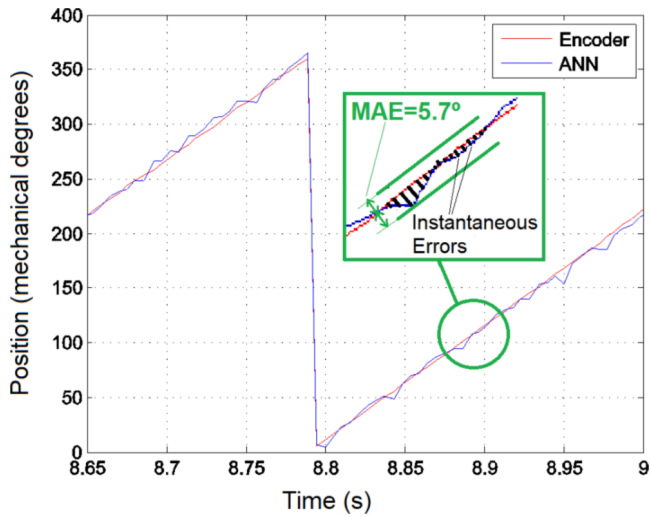

**Fig. 11.** Analysis of position estimation error when the ANN-based method tracks the encoder reference for a motor speed of 175 rpm.

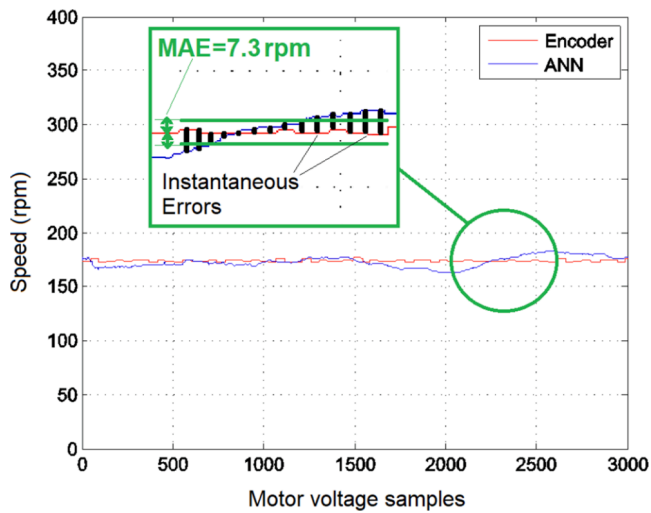

**Fig. 12.** Analysis of speed estimation error when the ANN-based method tracks the encoder reference for a motor speed of 175 rpm.

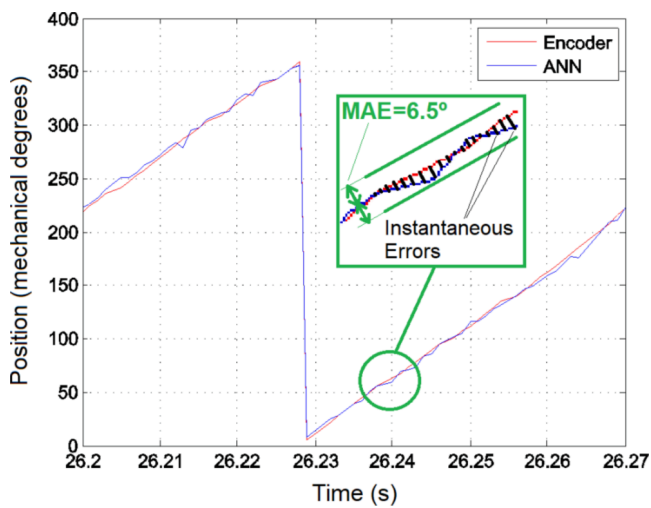

**Fig. 13.** Analysis of position estimation error when the ANN-based method tracks the encoder reference for a motor speed of 850 rpm.

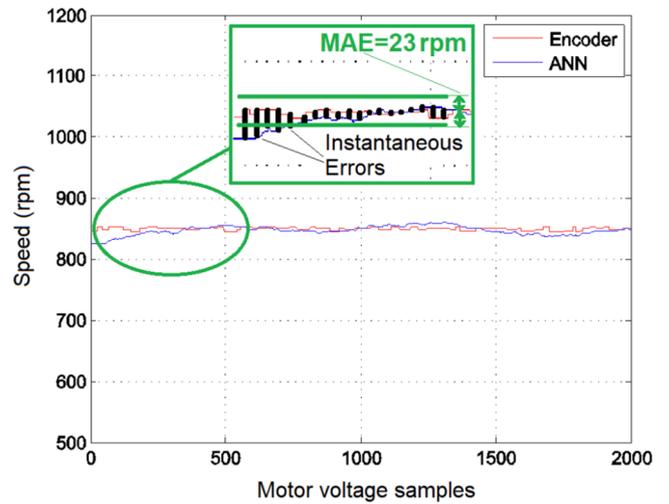

**Fig. 14.** Analysis of speed estimation error when the ANN-based method tracks the encoder reference for a motor speed of 850 rpm.

### 4.4. Comparison to related research

The comparison of the proposed method with related research works is discussed in detail below. The position and speed errors obtained by conventional and advanced methods are considered to compare their performance with respect to the method proposed in the present work, taking into account similar test conditions as far as possible. The related research works considered in the study are included in Tables 7 and 8. In these tables, "N/A" means Not Available and "-" indicates that a specific data is not described in the associated reference.

Firstly, the comparison of the position estimation performance of the proposed method with respect to the conventional methods is analysed. It is found that although the present work obtains an absolute position error of 0.8° (electrical degree) in the full range of 125–1,500 rpm, some of the most relevant conventional methods obtain errors of 60° [11,28], 30° [9,11,12,48,49], 3° [50], 3.5° [51], 10° [51], 5° [51], 10° [29], and 3.5° [52]. These data show that the proposed method provides an average error reduction of 95.1% and an error reduction of 73.3% with respect to the best conventional method. It is relevant to analyse some characteristics of the conventional methods that contribute to their limited performance:

- The more conventional sensorless method [11,28] obtains a position error of 60° because the voltage vectors used in the PWM inverter are only restricted to the six basic nonzero states of the switches. In the proposed method, not only are the six basic states considered, but phase transitions are also included in the VSNs to provide more states to train the ANNs.
- The methods based on BEMF signals and some improvements [9,48,49] obtain a position error of 30° in the transient state since the estimated commutation points are shifted with this value at the zero-crossing. Furthermore, an improvement that compensates for the ZCD error of the phase voltage achieves a position error of 3° [50]. The proposed method improves these results, as the ANNs act as a data filter to remove outliers after phase commutation that could influence on the position detection. To improve the BEMF-based position detection, the SMO method [29] is usually applied to remove multi-order harmonics from the BEMF signals. To limit the influence of harmonics in the proposed method, the signal processing is carried out with amplifiers and filters in the conditioning stage.

Secondly, the comparison of the position estimation performance of the proposed method (0.8° error) with respect to the advanced methods is analysed. Some of the most relevant advanced methods obtain errors

**Table 7**
Comparison performance of the present work to the related research.

| Ref. | Method | Position Error (electrical degree) | Speed Error (rpm) | Test Conditions |
|------|--------|-----------------------------------|-------------------|-----------------|
| | **Present work** | **< 0.8° (6.5° mech.)** | **< 22 rpm (5%)** | Experimental setup: 16-pole motor, speed 125–1,500 rpm, medium rated load 35 mNm. |
| [11,28] | Conventional method | 60° | – | Typical setup: PWM inverter restricted to six non-zero states of the switches. |
| [9,12] | BEMF ZCD: Typical | 30° | ~5 rpm (0.16%) | Simulation: 6-pole/8-pole motor, speed 3,000 rpm, rated load 2.1 Nm (in [9]). |
| [48] | BEMF ZCD: Improvement (inverter PWM duty adjusted to speed, wide speed range) | 30° | – | Experimental setup: 5-pole motor, speed 200–3,000 rpm, rated load 0.18 Nm. |
| [49] | BEMF ZCD: Improvement (BEMF observer, zero-crossing detection not required) | 30° | < 5 rpm (0.01%– 0.33%) | Simulation: 4-pole motor, speed 20–1,500 rpm, zero to rated load 1.5 Nm. |
| [50] | BEMF ZCD: Improvement (two-stage error compensation of line-voltages zero-crossing) | 3° | – | Experimental setup: 6-pole motor, speed 5,000–10,000 rpm, load 15 mNm. |
| [11] | Current variation with the magnetic saturation of the stator core | 30° | – | Simulation (aimed to initial position from standstill): 8-pole motor, speed 7,200 rpm, load N/A. |
| [51] | BEMF Integration | 3.6° (simulation) 3.5° (experimental) | – | Setup: 4-pole motor, speed 3,000 rpm (simulation) and 1,500 rpm (experimental), medium rated load 55 mNm. |
| | BEMF Third Harmonic | 3.6° (simulation) 10° (experimental) | – | |
| | Terminal Voltage Sensing | 3.6° (simulation) 5° (experimental) | – | |
| [29] | SMO: Typical | 10° | < 80 rpm | Experimental setup: 6-pole motor, speed 900–3,000 rpm, load 25 mNm. |
| [29] | SMO: Improvement (Adaptive) | 10° | < 30 rpm | |
| [52] | SMO: Improvement (2nd Order Integrator) | < 3.5° | – | Experimental setup: 10-pole motor, speed 90–1,200 rpm, 20–80% rated load (N/A). |
| [53] | SMO: Improvement (2nd Order Integrator and Tracking Differentiator) | 1.8° | 180 rpm (0.85%) | Experimental setup: 2-pole motor, high speed 21,000 rpm, no load. |
| [54,55] | Flux-Linkage function | < 1.75° (20–100 rpm) < 1° | – | Experimental setup: 4-pole motor, speed |

**Table 7** (*continued*)

| Ref. | Method | Position Error (electrical degree) | Speed Error (rpm) | Test Conditions |
|------|--------|-----------------------------------|-------------------|-----------------|
| | | (100–600/900 rpm) | | 20–600/900 rpm, load range 0.5–3.2 Nm. |

**Table 8**
Comparison performance of the present work to the related research (Continuation).

| Ref. | Method | Position Error (electrical degree) | Speed Error (rpm) | Test Conditions |
|------|--------|-----------------------------------|-------------------|-----------------|
| | **Present work** | **< 0.8° (6.5° mech.)** | **< 22 rpm (5%)** | Experimental setup: 16-pole motor, speed 125–1,500 rpm, medium rated load 35 mNm. |
| [56] | EKF algorithm | 1.8° | ~15 rpm | Simulation: 4-pole motor, speed 3,700 rpm, rated load 0.89 Nm |
| | | 2° | ~17 rpm | Simulation: 4-pole motor, speed 3,825 rpm, medium rated load 0.45 Nm. |
| [56] | Smoothing Filter algorithm | 0.95° | ~10 rpm | Simulation: 4-pole motor, speed 3,700 rpm, rated load 0.89 Nm. |
| | | 1° | ~12 rpm | Simulation: 4-pole motor, speed 3,825 rpm, load 0.45 Nm. |
| [57] | Virtual Hall Signals with Low-Pass Filters and Phase Shifters elimination | 3.5° (simulation) 4° (experimental) | – | Simulation and experimental setup: 2-pole motor, high speed 10,000 and 15,000 rpm, medium rated load. |
| [38] | Derivative of the Terminal Phase Voltages | 1.25°–3.75° | ~3 rpm | Experimental setup: 16-pole motor, speed 5–1,500 rpm, no load. |
| | | 1.25°–1.9° | ~1 rpm | Experimental setup: 16-pole motor, speed 5–150 rpm, rated load 53 mNm. |
| [33] | Double ANN topology (current and BEMF models) | 2.5° | – | Experimental setup: speed 800 rpm, load 0.5 Nm. |
| [34] | Feed-forward ANN topology | 8.3° (low speed) 3.5°, 4.9° (medium speeds) 12.8°, 18.9° (high speeds) | – | Experimental setup: 8-pole motor, low speed (1,000 rpm), medium speeds (2,000 rpm, 4000 rpm), high speeds (6,000 rpm, 8,000 rpm), load 50 Nm. |

of 1° [54,55], 1.8° [56], 0.95° [56], a range of 1.25° to 3.75° [38], 2.5° [33], and a range of 3.5° to 8.3° [34]. These data show that the proposed method provides an average error reduction of 67.2% and an error reduction of 20% with respect to the best advanced method. Although some of the advanced methods have a similar performance to that proposed, their design is more complex and requires more computational resources as explained below:

- The Smoothing Filter and EKF algorithms [56] obtain position errors smaller than 2° when considering the covariance matrices in the estimation, which means using hardware with a relatively high computing capacity to solve the operations of sum and products efficiently. The proposed method uses some of the simplest ANN training algorithms, such as the Backpropagation algorithm, to facilitate implementation in hardware with computational constraints.

- The Flux-Linkage function [54,55] obtains a position error of 1° in a range relatively far from standstill, but the function is calculated by integrating the motor voltage equation over an extended period of time. It requires a relative great amount of memory and specific computational resources to perform the integration operations (sums and products), which are generally performed with mathematical coprocessors that are not available in low-cost commercial hardware. The ANN algorithms used in the proposed method are trained with a reduced number of examples to minimise the memory usage, but a conditioning stage is required to obtain motor data with a low amount of spurious components.

Thirdly, the comparison of the speed estimation performance of the proposed method (22 rpm error in the full range) with respect to the conventional methods is analysed taking into account only the related research that provides numerical data of the speed errors. Some of the most relevant conventional methods are the BEMF Zero-crossing detection with a typical implementation and an improvement with a BEMF observer that obtains speed errors of 5 rpm [9,12,49], the SMO that obtains a speed error of 80 rpm with a typical implementation [29], and the adaptive SMO that obtains an error of 30 rpm [29]. These data show that the proposed method provides an average error reduction of 26.7% and with respect to the best conventional method, an error increment of 340%.

Finally, the comparison of the speed estimation performance of the proposed method with respect to the advanced methods is analysed, and, as indicated previously, only the related research that provide numerical data are taken into account. Some of the most relevant advanced methods are the EKF algorithm that obtains a speed error of 15 rpm [56], the Smoothing Filter algorithm that obtains an error of 10 rpm [56], and the Derivative of the Terminal Phase Voltages that obtains an error between 1 rpm and 3 rpm [38]. These data show that the proposed method provides an average error increment of 144.4% and with respect to the best advanced method, an error increment of 1000%. It is relevant to analyse some characteristics of the advanced methods with respect to the present work:

- The Derivative of the Terminal Phase Voltages [38] obtains a speed error <3 rpm. Its performance is based on the implementation of algorithms hardcoded with specific hardware and embedded software, such as conditioning stages for signal amplification and derivation, and ghost-spike filters for derivation pulses. The disadvantage of this design is that the effort to implement hardware for in-circuit processing is greater than the required in the proposed method to deploy simple ANN algorithms in commercial devices such as FPGAs.
- Other methods that use ANNs, such as a double ANN topology [33] and feed-forward topology [34], do not provide numerical results of speed errors, as they are usually aimed to estimate position. Their position errors are between 2.5° and 8.3° in the equivalent range of low and high speeds, which represents an average error increment of 69.8% regarding the proposed method. If these results were extrapolated to the speed estimation, their speed error would be hypothetically greater than that obtained in the present work.

The comparison of the present work with other methods has limitations related to the test conditions. As far as possible, similar test conditions based on the proposed method are considered, such as speed ranges close to 125–1,500 rpm (far from standstill) and results from experimental setups rather than computer simulations. Another detail to consider is the number of poles of the motors used in the tests, which have at least four poles and motors with fewer poles are usually for speeds outside the target range. For reference only, Tables 7 and 8 include two methods tested on 2-pole motors for speeds above 10,000 rpm [53,57], which also obtain greater position errors than in the proposed method.

In summary, the proposed method significantly improves the average performance of the position estimation over conventional and advanced methods, surpassing even the most remarkable methods analysed in each category. Furthermore, the proposed method achieves a slightly better average performance of the speed estimation over conventional methods, but obtains a worse overall performance of the speed estimation over the advanced ones. These results show that a possible refinement of the method can be considered in the speed estimation based on the methods discussed above, such as the BEMF observer that obtains a significant reduction of the error to 5 rpm [49] with moderate complexity to be applied in precision and critical applications. If the BEMF observer method was considered as the basis for improving the proposed method, the motor BEMF signals should be acquired to feed the inputs of the speed estimation ANN.

## 5. Conclusion

In this article, a new sensorless method is proposed for position and speed estimation of BLDC motors using only the terminal phase voltages. The method innovates in considering phase transitions as virtual extended rotor steps to improve the rotational angle resolution in ANN training, uses a multilayer perceptron-based ANN topology with a non-complex training algorithm that requires low computational requirements such as memory and processing capacity, allowing it to be implemented in a low-cost FPGA, and only needs elemental signal processing based on amplification and filtering to attenuate spurious and harmonic components from motor signals, enabling feature synthesis with high signal to noise ratio. Regarding related research works, the results show the following:

- Compared to conventional methods such as the BEMF Zero-crossing detection and improvements based on BEMF observers, the Terminal Voltage Sensing, and the SMO, the proposed method provides an average error reduction of 95.1% for the position estimation, and an error reduction of 73.3% with respect to the best conventional method. In addition, for the speed estimation, it obtains an average error reduction of 26.7%, but the error increases by 340% regarding the best conventional method.
- Compared to advanced methods such as the Derivative of the Terminal Phase Voltages, the EKF algorithm, the Smoothing algorithm, and algorithms with similar ANN topologies, the proposed method provides an average error reduction of 67.2% for the position estimation, and an error reduction of 20% with respect to the best advanced method. In addition, for the speed estimation, it obtains an average error increment of 144.4% and an error increment of 1000% regarding the best advanced method.

In conclusion, this study presents a simple method with low computational requirements, which significantly improves the overall performance of the position estimation over conventional and advanced methods. Also, the overall performance of the speed estimation is slightly improved over conventional methods, but worse speed estimation is achieved than the advanced ones.

*CRediT authorship contribution statement*

**Jose-Carlos Gamazo-Real:** Project administration, Funding acquisition, Conceptualization, Supervision, Investigation, Methodology, Resources, Software, Formal analysis, Validation, Data curation, Visualization, Writing – original draft, Writing – review & editing. **Víctor Martínez-Martínez:** Investigation, Visualization, Formal analysis, Validation, Data curation, Writing – original draft. **Jaime Gomez-Gil:**

Resources, Writing – review & editing.

**Declaration of Competing Interest**

The authors declare that they have no known competing financial interests or personal relationships that could have appeared to influence the work reported in this paper.

**Acknowledgements**

Funding for open access charge: Universidad Politécnica de Madrid, Spain. In addition, this work was partially supported by the Comunidad de Madrid under Convenio Plurianual with the Universidad Politécnica de Madrid in the actuation line of Programa de Excelencia para el Profesorado Universitario.